\newcommand{\nn}{\nonumber}
\newcommand{\eps}{\varepsilon}
\newcommand{\fig}{\figurename~}

\documentclass{elsart}
\usepackage{psfrag}
\usepackage{latexsym}
\usepackage{graphicx}
\usepackage{color}

\journal{Solid State Communications}
\begin{document}
\begin{frontmatter}

\def\runtitle{
Dynamical Mean Field Theory equations on nearly real frequency axis}

\def\runauthor{Fathi, Jafari}

\title{Dynamical Mean Field Theory equations on nearly real frequency axis}

\author[sut]{M. B. \textsc{Fathi}}
\ead{mb.fathi@gmail.com}
\author[iut]{S. A. \textsc{Jafari}}
\ead{sa.jafari@cc.iut.ac.ir}

\address[sut]{Department of Physics, Sharif University of Technology, Tehran 14588-89694, Iran}
\address[iut]{Department of Physics, Isfahan University of Technology, Isfahan 84156, Iran}

\begin{abstract}
\noindent The Iterated Perturbation Theory (IPT) equations of
the Dynamical Mean Field Theory (DMFT) for the half-filled Hubbard
model, are solved on nearly real frequencies at various values of
the Hubbard parameters, $U$, to investigate the nature of
metal-insulator transition (MIT) at finite temperatures. This method
avoids the instabilities associated with the infamous Pad\'e
analytic continuation and reveals fine structures across the MIT at
finite temperatures, which {\em can not be captured} by conventional
methods for solving DMFT-IPT equations on Matsubara frequencies. Our
method suggests that at finite temperatures, there is an abrupt
decrease in the height of the quasiparticle (Kondo) peak at a
critical value of $U_c$, to a non-zero, but small bump, which
gradually suppresses as one moves deeper into the {\em bad}
insulator regime. In contrast to Vollhardt \& coworkers [J. Phys.
Soc. Jpn. {\bf 74}, 136, 2005] down to $T=0.01$ of the
half-bandwidth we find no $T^*$ separating bad insulator from a true
Mott insulator.\\

\end{abstract}

\end{frontmatter}

\section{Introduction}
  Feynman introduced a graphical representation of various
perturbation schemes in the field theory which bears his
name\cite{Flensberg}, Feynman diagrams. In condensed matter
physics, the natural high momentum cut-off provided by the lattice
dimensions prevents ultraviolet divergences. But yet remains the
practical task of calculating them and interpreting the results.

  Matsubara on the other hand introduced a very clever
mathematical trick, which not only combines the thermal and quantum
mechanical averaging required in calculations of the correlation
functions in a neat way and treats them on the same footing, but
also provides a very convenient method for numerical calculations of
the diagrams. The essential ingredient is that the Wick rotation
$\tau=it$ replaces the oscillatory $e^{-i\xi_kt}$ factors by
decaying $e^{-\xi_k\tau}$ factors in imaginary time, which are very
convenient for putting on computer and convergence is particularly
fast in iterative or self-consisted formulations of perturbation
theory. However, the price will be payed when one has to undo the
Wick rotation at the end of calculation to obtain dynamical
quantities by replacing $i\omega_n\to \omega+i\eta$, where $\eta$ is
an infinitesimal positive constant.

   The hurdle one faces in undoing the Wick rotation is
that, if using the Pad\'e approximation one fits a quotient of two
polynomials $f_N(z)$ and $g_M(z)$ to the table of data obtained for
Matsubara frequencies $i\omega_n$, and then replaces
$z\to\omega+i\eta$ in the resulting function, the calculated
spectral weight are not always stable with respect to variations in
parameters $N,M$. Even if for some parameter regime, or for some
particular problem one obtains relatively stable results, the
aforementioned disrepute of the Pad\'e approximation warns us about
the reliability and/or the quality of the dynamical quantities
obtained in this way.

People have been worried about this question from time to time, and
there has been some proposals for reliable way of using the Pad\'e
approximation in analytic continuation of numerical data: Beach and
coworkers~\cite{Beach} proposed a symbolic computer aided algebra
with arbitrary precision (typically 100-200 decimal places, which
lack in single or double precision arithmetics of standard
programming languages like \verb|C++| or \verb|Fortran|). They also proposed a
qualitative measure of the reliability of continued data. Mishchenko
and collaborators~\cite{Mishchenko2000,Mishchenko2002} proposed an
stochastic optimization method which allows one to handle both broad
and sharp features of the spectrum on equal footing.

On the other hands, Schmalian and colleagues proposed an alternative method which is
quite intuitive and general\cite{Schmalian}: Instead of solving the
diagrammatic equations for Matsubara frequencies $i\omega_n$, solve
them for frequencies $\omega+i\gamma$, where $\gamma$ is a finite
constant. The {\em finite} value of $\gamma$ (usually taken to be less
than the first nonzero Matsubara frequency) is important, and
provides the damping required for convergence of the iterative
solutions. At the same time analytic continuation from
$\omega+i\gamma$ to $\omega+i\eta$, where $\eta=0^+$, not only
is stable, but also sustains fine features of the spectral function,
such as shadow bands of the high temperature
superconductors\cite{SchmalianPRL}, as they found by applying this
method to solve the diagrammatic equations of the fluctuation
exchange approximation\cite{Bickers}.

Dynamical mean field theory approximation has been successful in
addressing the issue of metal-insulator transitions in correlated
electron systems at zero temperature\cite{GeorgesRMP}. Keeping the aforementioned
concerns about the reliability of the Pad\'e analytic continuation
procedure in mind, in this paper we use the method of Schmalian
{\em et. al}\cite{Schmalian} to re-examine the nature of MIT in the
half-filled Hubbard model at finite temperatures. We find that
paying the price at the beginning and solving slightly more
difficult equations for $\omega+i\gamma$ pays off and in addition to
providing reliable Pad\'e analytic continuation with absolutely
no negative spectral weights, reveals fine
structure in the insulating side of the MIT. A small bump in the
spectral weight which persists in the insulating phase, is
stably produced in our approach and can not be captured by Pad\'e
analytical continuation of the solutions of DMFT equations for
Matsubara frequencies.

   The only difficulty with this method is that with conventional
double precision numeric accuracy  at lower temperatures it is hard
to achieve convergence. But the asymptotic behaviors at $T\to 0$
limit in our approach agree with other methods of solving the DMFT
equations.

The paper is organized as follows: First we analytically continue
the IPT equations of DMFT to $\omega+i\gamma$ line above the real
frequency axis. Then we present the numerical solutions of the
resulting equations for various values of the Hubbard parameters, $U$,
at half-filling, and elevated temperatures. Finally we present our
conclusions.

\section{Formulation}
Within DMFT approximation, the problem of interacting electrons on a
lattice can be mapped onto an effective impurity problem surrounded
by a self-consistent bath. The impurity Green's function, $\cal G$,
is related to it's bare counterpart via the Dyson
equation\cite{GeorgesRMP},
\begin{eqnarray}
   {\cal G}_0^{-1}=\Sigma+\frac{1}{\tilde{D}(i\omega_n+\mu-\Sigma)}
   \label{eq:dmft}
\end{eqnarray}
where,
\begin{eqnarray}
   \tilde{D}(i\omega_n+\mu-\Sigma)=\int_{-\infty}^{\infty}
   {d\eps\:\frac{D(\eps)}{i\omega_n+\mu-\Sigma-\eps}}
   \label{eq:hilb}
\end{eqnarray}
is the Hilbert transform of density of states~(DOS).
In (\ref{eq:hilb}), $\tilde{D}(i\omega_n+\mu-\Sigma)$ is the on-site
full Green's function for site $o$ \emph{i.e.} $G_{oo}$ and it's
imaginary part gives the interacting DOS,
\begin{equation}
   G(i\omega_n)=\tilde{D}(i\omega_n+\mu-\Sigma(i\omega_n))\label{eq:onsiteG}
\end{equation}

In IPT approximation, the self-energy is given by the second order perturbation
theory\cite{GeorgesRMP} as,
\begin{equation}
   \Sigma(i\omega_n)\simeq
   \underbrace{U^2\int_{0}^{\beta}{d\tau\:e^{i\omega_n\tau}\hat{\cal{G}}_0(\tau)^3}}_{\Sigma^{(2)}}.
   \label{eq:ipt}
\end{equation}

DMFT equations written in Matsubara form yield no dynamical
quantities, until the analytical continuation to real frequency axis
is done,
\[i\omega_n\rightarrow\omega+i\eta\]
To see where lies the root of numerical problems, one notes that
the real-frequency and imaginary time Green's functions
are connected by~\cite{Dopf}
\begin{equation}
G(\tau)=\frac{1}{\pi}\int_{-\infty}^{\infty}{d\omega}\:\frac{e^{-\tau\omega}}{1+e^{-\omega/T}}
Im G(\omega+i0^+) \label{eq:analcontin}
\end{equation}
where, $G(\tau)=T\sum_n e^{i\omega_n\tau}G(i\omega_n)$ is the
Fourier transform of Matsubara function.
Due to exponential factors, the small changes
in $G(\tau)$ (equivalently in $G(i\omega_n)$) are associated
with large changes in $G(\omega+i\eta)$.

Now we turn our attention to the question of analytic continuation
of DMFT equations in IPT approximation, parallel to the work of
Schmalian and coworkers\cite{Schmalian}.
We rewrite the equation to be solved for nearly real frequencies
$\omega+i\gamma$, with a finite $\gamma$.
Then, we go to the limit $\gamma\to 0^+$  via Pad\'e approximation.
Pad\'e approximation at this stage turns out to be stable.
The finite parameter $\gamma$ is chosen to provide the attenuation
factors (as will be seen below) needed for convergence of the
self-consistent equations.

The first equation to be continued analytically to nearly real
axis is (\ref{eq:onsiteG}), which bears no
difficulty,
\begin{equation}
G(\omega+i\gamma)=\tilde{D}(\omega+i\gamma+\mu-\Sigma(\omega+i\gamma)).
\end{equation}

The next equation to be continued to nearly real-frequency axis is
(\ref{eq:dmft}), which again simply reads
\begin{equation}
{\cal G}_0^{-1}(\omega+i\gamma)=\Sigma(\omega+i\gamma)
+\frac{1}{\tilde{D}(\omega+i\gamma+\mu-\Sigma(\omega+i\gamma))}
\end{equation}

The main problem relies on analytically continuing the IPT approximation,
Eq.~(\ref{eq:ipt}) to nearly real frequency axis. This equation
depends on the frequency not only through the Fourier factor
$e^{i\omega_n \tau}$, but also through the Green's function ${\cal
G}_0(\tau)$. In this case, analytical continuation must be performed
through the change of integral to it's retarded form. Here the
difficulty arises from the fact that we've solved the action
equation (for a detailed review see Georges \emph{et
al.}\cite{GeorgesRMP}, section III.A), on the imaginary axis to
yield Matsubara function, so if one wishes to gain the physical
quantities (such as retarded green's function) one must either solve
the problem originally on real frequency axis or analytically
continue it. Kajueter and Kotliar\cite{kajueter, held} made an ansatz
for self-energy on the real frequency axis of the form
\begin{equation}
\Sigma(\omega)=U+\frac{A\Sigma^{(2)}(\omega)}{1-B\Sigma^{(2)}(\omega)}
\label{eq:ipt-realfreq}
\end{equation}
where, $\Sigma^{(2)}$ is the second order contribution to
self-energy from (\ref{eq:ipt}). Of course, this equation alone
doesn't solve the problem, since there are other functions written
in Matsubara form. Fortunately there is another way to overcome the
problem. Equation (\ref{eq:ipt}) can be written as
\begin{equation}
\Sigma(i\omega_n)\simeq
U^2\int_{0}^{\beta}{d\tau\:e^{i\omega_n\tau}\hat{\cal{G}}_0(\tau)\hat{\cal{G}}_0(\tau)\hat{\cal{G}}_0(-\tau)}.
\end{equation}
Using the Fourier transformation
\begin{equation}
{\cal G}_0(\tau)=\sum_{n=-\infty}^{\infty}{e^{i\omega_n\tau}{\cal G}_0(i\omega_n)},
\end{equation}
it can be written as
\begin{equation}
\int_{0}^{\beta}\!\!\!{d\tau\:e^{i\omega_n\tau}{\cal G}_0(\tau){\cal
G}_0(\tau){\cal G}_0(-\tau)}=\sum_k{\cal
G}_0(i\omega_k)\chi^0(i(\omega_n+\omega_k)),
\end{equation}
where,
\begin{eqnarray}
\chi^0(i(\omega_n+\omega_k))=\sum_l{\cal G}_0(i\omega_l){\cal
G}_0(i(\omega_n+\omega_l+\omega_k))\label{eq:chi}
\end{eqnarray}
is the particle-hole bubble.

With the aid of contour integration and employing the complex forms of Fermi and Bose
functions,
\begin{equation}
f(z)=\frac{1}{e^{\beta z}+1},~~~n(z)=\frac{1}{e^{\beta z}-1}
\end{equation}
which have their poles exactly at Matsubara frequencies
$z=i\omega_n=i(2n+1)\pi/\beta$ and
$z=i\omega'_n=i(2n)\pi/\beta$, respectively, the summation over imaginary frequencies can be done,
\begin{eqnarray}
\chi^0(i\nu_k)&=&\sum_l{\cal G}_0(i(\nu_n+\omega_l)){\cal G}_0(i\omega_l)\nonumber\\
              &=&\int_{-\infty}^{\infty}{\frac{d\eps}{2\pi i}\:}\\
&&\big\{f(\eps+i\gamma'){\cal G}_0(i\nu_k+\eps+i\gamma'){\cal
G}_0(\eps+i\gamma')\nn\\
&-&f(\eps-i\gamma'){\cal G}_0(i\nu_k+\eps-i\gamma'){\cal
G}_0(\eps+i\gamma')\nn\\
&+&f(\eps+i\gamma'){\cal G}_0(\eps+i\gamma'){\cal
G}_0(\eps+i\gamma'-i\nu_k)\nn\\
&-&f(\eps-i\gamma'){\cal G}_0(\eps-i\gamma'){\cal
G}_0(\eps-i\gamma'-i\nu_k) \big\}.\nn
\end{eqnarray}
Now, performing the analytical continuation
$i\nu_k\rightarrow\omega+i\gamma$, and making use of Kramers-Kronig
transformation for the green function, along with the Laplace transformation
\begin{equation}
\frac{1}{\omega+i\gamma}=-i\int_{0}^{\infty}{dt~e^{i(\omega+i\gamma)t}},
\end{equation}
the particle-hole bubble reduces to
\begin{equation}
\chi^0(\omega+i\gamma)=\int_{0}^{\infty}{dt\:\chi^0(t)e^{i(\omega+i\gamma)t}},
\end{equation}
where,
\begin{eqnarray}
\chi(t)=-i(2\pi)^2\left(\rho(t)\left[A^\ast(t)+A(-t)e^{2\gamma
t}\right]\right.\nn\\
\left.-\rho^\ast(t)\left[A(t)+A^\ast(-t)e^{2\gamma t} \right]\right)
\label{chi0.eqn}
\end{eqnarray}
In the above formula the spectral densities are given by,
\begin{equation}
\rho(t)=-\frac{1}{2\pi}\frac{1}{\pi}\int_{-\infty}^{\infty}{d\eps\:\textit{Im}{\cal G}_0(\eps+i\gamma)
e^{-i\eps t}},
\label{rhot.eqn}
\end{equation}
and,
\begin{equation}
A(t)=\frac{i}{2\pi}\int_{-\infty}^{\infty} {d\eps\:
f^\ast(\eps+i\gamma){\cal G}_0^\ast(\eps+i\gamma)}e^{-i\eps t}.
\label{At.eqn}
\end{equation}
In an analogous manner, the Laplace transform of self-energy
can be written as:
\begin{eqnarray}
\label{Sigt.eqn}
\Sigma(t)&=&-i2\pi T\:U^2\:Re\chi^0(0+i0^+)\:e^{\gamma
t}\rho(t)\nn\\
&&+i(2\pi)^2U^2\times\left(v^0(t)[A(t)+A^\ast(-t) e^{2\gamma t}]\right.\nn\\
&&~~~~~~~~~~~~~~\left.-\rho(t)[B^\ast(t)+B(-t)e^{2\gamma t}]\right)
\label{eq:selfenergyoneareal}
\end{eqnarray}
where,
\begin{eqnarray}
v^0(t)=-\frac{1}{2\pi}\frac{1}{\pi}\int_{-\infty}^{\infty}{d\eps\:\textit{Im}\chi^0(\eps+i\gamma)
e^{-i\eps t}},
\label{v0.eqn}
\end{eqnarray}
and
\begin{eqnarray}
B(t)=\frac{i}{2\pi}\int_{-\infty}^{\infty} {d\eps\:
n^\ast(\eps+i\gamma)\chi^{0\ast}(\eps+i\gamma)}e^{-i\eps t}.
\label{B.eqn}
\end{eqnarray}
\section{Numerical Results}
The numerical results are given in units of the half bandwidth
$W/2=1$. The temperature $T=0.048$ and the Hubbard parameter $U$
varies from $U=0$ through $7.5$.

To perform the Fourier transforms we used the FFT routines. To
overcome the alias effects arising from the $1/\omega$ tail, we
chose the energy mesh large enough to ensure that the function vanishes
at the boundaries. We have used $N=2^{15}$ frequency points in
an energy range of $2\omega_{max}=250$. Also we have taken
$\gamma=\pi T/4$. The finite value of $\gamma$ provides necessary
damping factors in the convergence of the iteration process.
Therefore at lower temperatures it becomes harder to achieve
convergence.  Convergence at lower temperatures requires
smaller values of $\gamma$ and hence higher accuracy than
conventional double precision~\cite{Beach}.

There are many lattices examined by the DMFT method, some of which
is mentioned in the review paper by Georges \emph{et al.}~\cite{GeorgesRMP}.
We work with Beth\'e lattice with semicircular DOS~\cite{Economou}.

We begin with an initial guess for the green function,
${\cal G}_0(\omega+i\gamma)$. Using Eqs.~(\ref{rhot.eqn},\ref{At.eqn}) we calculate $\rho(t)$
and $A(t)$ which are used to evaluate $\chi^0(t)$ in Eq.~(\ref{chi0.eqn}).
Then we calculate $v^0(t)$ and $B(t)$, Eqs.~(\ref{v0.eqn},\ref{B.eqn}),
through which the self-energy $\Sigma(t)$, Eq.~(\ref{Sigt.eqn}),
and hence its Laplace transform, $\Sigma(\omega+i\gamma)$, can be
evaluated. To close the iteration loop,
$\Sigma(\omega+i\gamma)$ is inserted in the self-consistency
equation (\ref{eq:dmft}) to update the initial Greens' function
and iteration is performed until convergence is achieved.

We have obtained converged solutions for $U$ in the range $[0,7.5]$
for $T=0.048$. At this temperature the transition occurs at critical
value $U_c\approx 3.0^+$, in agreement with the results of Zhang and
coworkers~\cite{XYZhang}, and Vollhardt et
al.~\cite{voll-appl-real-mat}.

\subsection{The spectral density before the transition}
\begin{figure}[htbp]
\centering
    \includegraphics[width=3in]{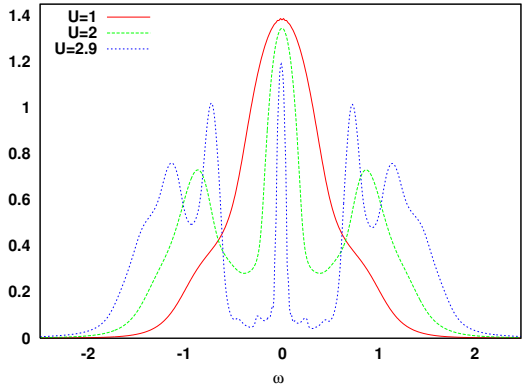}
    \caption{DOS at $T=0.048$ and values of $U<U_c$ in the
    metallic side. When the Hubbard parameter is increased the weight of fine
    peak is transferred to it's neighbors. unlike the zero
    temperature case, the height of the central Kondo peak
    is not constant.
    \label{fig.DOS-before-tran}}
\end{figure}
Before the transition in the metallic regime, as one increases $U$ the DOS tends to
split, leaving a central Kondo peak corresponding to quasiparticle
at the Fermi level (\fig\ref{fig.DOS-before-tran}) in agreement with the conventional methods of solving the
DMFT equations~\cite{GeorgesRMP}.
In our computer code we have not forced the
height of the Kondo peak to be constant.
In our finite $T$ solution unlike the zero temperature case
the hight of the Kondo resonance is not constant during the
transition~\cite{muller}.
The height of the quasiparticle peak at Fermi energy is instead gradually
redistributed and shifted to the upper(lower) edge of the
lower(upper) Hubbard band.

\subsection{The spectral density near the transition}
\begin{figure}[htbp]
    \centering
  \includegraphics[width=3in]{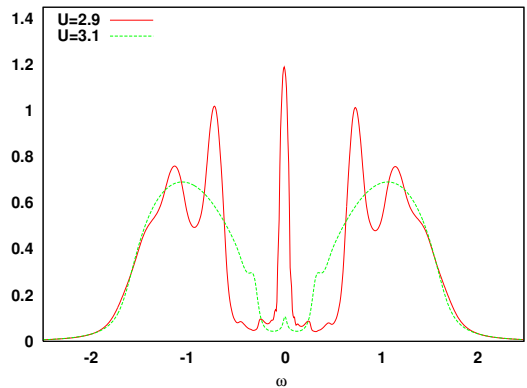}\\
  \caption{The evolution of DOS when the transition occurs.
  The results is shown for two representative values
  $U=2.9$ and $U=3.1$ at temperature $T=0.048$.}\label{fig.atthe-tran}
\end{figure}

The transition occurs at a critical Hubbard parameter $U_c$. We
find the value for this critical Hubbard parameter in the
range $2.99\leq U_c\leq 3.01$. At the transition  the height
of peak falls off suddenly to a {\em very small, but nonzero value},
\fig\ref{fig.atthe-tran}.

\subsection{The spectral density after the transition}
\begin{figure}[btbp]
\centering
    \includegraphics[width=3in]{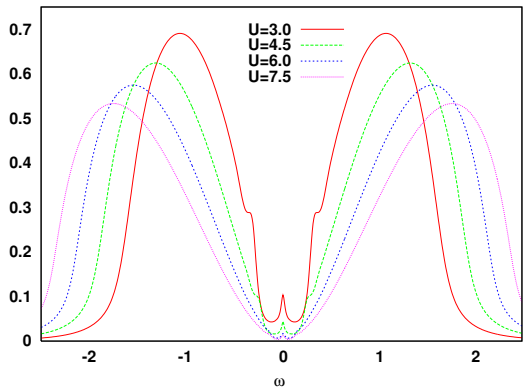}
    \caption{DOS after the transition, at temperature $T=0.048$.
    Increasing $U$ tends to push upper and lower Hubbard bands
    far apart, but the little peak resits and remains there at
    even much larger values of $U$.
    \label{fig.DOS-afterthe-tran}}
\end{figure}

The evolution of the spectral function after the transition is
shown in \fig\ref{fig.DOS-afterthe-tran}.
In our solution increasing $U$ results in depletion of the spectral
weight but does not give a clean gap at finite temperatures.
For larger values of $U$ the spectral density at Fermi level is
still finite and vanishes only in the limit $U\rightarrow \infty$,
so that there is no real transition but a crossover from a
metallic-like to an insulating-like solution. Vollhardt \emph{et
al.}\cite{voll-appl-real-mat} named this state as a {\em bad insulator}.

  The finite temperature analogue of the Mott-Hubbard gap can be
thought of as the peak-to-peak distance between the upper and lower Hubbard
bands in Fig.~\ref{fig.DOS-afterthe-tran}. As can be seen this
distance increases proportional to $U$.
With increasing $U$ the "gap" tends to
expand and the spectral weight in this region tends to decrease,
but as we mentioned the gap  is not clean at $T=0.048$. At
$U\to\infty$ one expects to obtain a clean gap around the Fermi
level. In order to obtain clean gap at finite large $U$
one has to go to $T\to 0$ limit.

\begin{figure}[htbp]
\centering
    \includegraphics[width=3in]{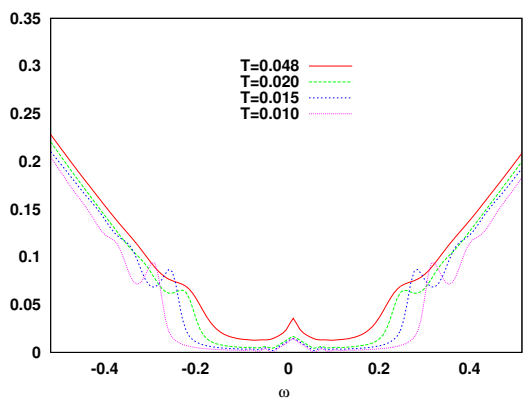}
    \caption{ The evolution of the spectral weights in the bad insulator regime for $U=5.0$
    for lower temperatures. The lower the temperature the cleaner the gap and the insulator
    evolves to Mott insulator. As the figure shows, the walls of the gap tends to vertical direction
    at $T\to 0$ limit.}
    \label{fig.dos-after-tran-T}
\end{figure}

In the solution obtained by Vollhardt \emph{et
al.}~\cite{voll-appl-real-mat} they introduce a temperature $T^*$
below which the bad insulator becomes a true insulator. Down to
$T=0.01$ where we have obtained converged solution in our method we
do not find such a $T^*$. The trend of the lower $T$ data in
Fig.~\ref{fig.dos-after-tran-T} shows that at $T\to 0$ limit the gap
starts to become cleaner and develops sharp edge characteristic of
IPT method~\cite{GeorgesRMP}.
The small central peak in the bad insulating side gradually
suppresses with decreasing $T$ but does not
vanish at any finite $U$ (Fig.~\ref{fig.dos-after-tran-T}).
The filling of the Mott-Hubbard gap by spectral
weight transfer with increasing temperature has recently been
detected experimentally by photoemission experiments\cite{mo}.

An important aspect of insulating phase obtained in the present method
is the presence of a small bump at the Fermi level which is
clearly seen in the Figs.~\ref{fig.DOS-afterthe-tran} and
\ref{fig.dos-after-tran-T}. This little peak
has a height about $10\%$ of neighboring bands and does not vanish after
the transition at any $U$.

\section{Summary and conclusions}
\label{sec:conc} We employed a new method for solving diagrammatic
equations to solve IPT equations of DMFT. We first analytically
continued the equations to nearly real frequencies. Solving the
continued equations is slightly harder, but seems to capture fine
features of the spectral weight not previously reported.

We found that the height of Kondo peak does not remain constant before transition
which is in contrast to the zero temperature result of M\"uller-Hartmann~\cite{muller},
which states that the height of the Kondo peak must remain constant
before the transition occurs.
The transition is first order and the Kondo peak suddenly falls off to a very small but
non-zero value.  For larger
values of $U$ this small spectral density at Fermi level will persist
and vanishes only in the limit $U\to \infty$ or
$T\to 0$. In this limit the gap also becomes cleaner and we obtain
a true insulator only in $T\to 0$ limit. Hence we do not find a
$T^*$~\cite{voll-appl-real-mat}.

\section*{Acknowledgement}
M.B.F. would likes to thank Prof. M.A. Vesaghi for his many
suggestions and constant support during this research. S.A.J. was
supported by ALAVI Group Ltd.

\end{document}